\documentclass[10pt,conference]{IEEEtran}
\IEEEoverridecommandlockouts
\usepackage{cite}
\usepackage{amsmath,amssymb,amsfonts}
\usepackage{algorithmic}
\usepackage{graphicx}
\usepackage{textcomp}
\usepackage{xcolor}
\usepackage{microtype}

\usepackage{multirow}
\usepackage{booktabs}
\usepackage{siunitx}
\usepackage{pifont}

\usepackage{soul}
\usepackage{hyperref}

\def\BibTeX{{\rm B\kern-.05em{\sc i\kern-.025em b}\kern-.08em
    T\kern-.1667em\lower.7ex\hbox{E}\kern-.125emX}}
    
\begin{document}

\title{Effective Pre-Training of Audio Transformers \\ for Sound Event Detection \\





\thanks{The computational results presented were achieved using the Linz Institute of Technology (LIT) AI Lab Cluster. The LIT AI Lab is supported by the Federal State of Upper Austria. Gerhard Widmer's work is supported by the European Research Council (ERC) under the European Union's Horizon 2020 research and innovation programme, grant agreement No 101019375 (Whither Music?).}}

\author{\IEEEauthorblockN{Florian Schmid$^1$, Tobias Morocutti$^1$, Francesco Foscarin$^{1,2}$, Jan Schlüter$^1$, Paul Primus$^1$, Gerhard Widmer$^{1,2}$}
\IEEEauthorblockA{\textit{$^1$Institute of Computational Perception, $^2$LIT Artificial Intelligence Lab} \\
\textit{Johannes Kepler University Linz, Austria}\\
{firstname.lastname}@jku.at}
}


\maketitle

\begin{abstract}
We propose a pre-training pipeline for audio spectrogram transformers for frame-level sound event detection tasks. On top of common pre-training steps, we add a meticulously designed training routine on AudioSet frame-level annotations. This includes a balanced sampler, aggressive data augmentation, and ensemble knowledge distillation. For five transformers, we obtain a substantial performance improvement over previously available checkpoints both on AudioSet frame-level predictions and on frame-level sound event detection downstream tasks, confirming our pipeline's effectiveness. We publish the resulting checkpoints that researchers can directly fine-tune to build high-performance models for sound event detection tasks.

\end{abstract}

\begin{IEEEkeywords}
Sound Event Detection,
Audio Transformers, Knowledge Distillation, AudioSet, Temporally-Strong Labels
\end{IEEEkeywords}

\section{Introduction}

Pre-Trained Audio Spectrogram Transformers have become prevalent across the audio domain, dominating audio classification tasks such as audio tagging~\cite{Gong21Ast,Koutini21Passt,Chen22HTS-AT,Liu23CAT,Gong22kd-ast,Dinkel24CED,Chen23BEATs,Niizumi24M2D,Niizumi24m2d-clap,Huang22Masked,Atito22ASiT,Gong2022ssast,Chong23maskspec} and sound event detection~\cite{li24atst-frame,Shao24atst-sed,schmid2024multi}. The former task addresses \textit{clip-level} predictions, i.e., labels associated globally with the entire audio clip. The latter is based on \textit{frame-level} labels, associated with specific time positions.
%
%
The largest and most widely used general-purpose audio dataset for pre-training audio transformers is AudioSet~\cite{audioset2017Gemmeke}. It contains both clip-level~\cite{audioset2017Gemmeke} and a subset with frame-level~\cite{audioset-strong} sound event annotations, usually referred to as \textit{AudioSet Weak} and \textit{AudioSet Strong}, respectively.

Many of the widely used pre-trained audio transformers employ a two-step pre-training pipeline: a first step, either self-supervised learning (SSL) on AudioSet~\cite{Dinkel24CED,Chen23BEATs,Niizumi24M2D,Niizumi24m2d-clap,Huang22Masked,Atito22ASiT,Gong2022ssast,Chong23maskspec} or supervised training on ImageNet~\cite{Gong21Ast,Koutini21Passt,Chen22HTS-AT}, followed by a supervised step on AudioSet Weak. 
While effective for clip-level tasks, the second step does not promote learning temporally precise representations and may be suboptimal for frame-level downstream tasks. 
Therefore, we propose to extend the pre-training pipeline with a supervised step on the strong annotations of AudioSet~\cite{audioset-strong}. 

AudioSet Strong has been used for varying purposes, including the evaluation of zero-shot models~\cite{li2024advancing} and streaming capabilities of transformers~\cite{dinkel2023streaming}, or to pre-train domain-specific components of text-to-audio grounding~\cite{xu2023investigating}, audio captioning~\cite{xie2023enhance} or audio generation~\cite{Zhang24FoleyCrafter} systems.
In \cite{li24atst-frame}, it is used as a downstream task to evaluate self-supervised pre-trained models.
However, to our knowledge, we are the first to train on AudioSet Strong as a step in a pre-training pipeline and rigorously evaluate its effectiveness across multiple transformer models and downstream tasks.





To make this step maximally effective, we carefully tune the pre-training strategy.
Our first proposal is to use a balanced sampler for the heavily imbalanced classes and apply aggressive data augmentation to avoid overfitting. This is particularly important, as AudioSet Strong ($\sim$100k) is considerably smaller than AudioSet Weak ($\sim$2M).
%
%
The second improvement is inspired by the performance gains of ensemble knowledge distillation (KD) techniques~\cite{allen-zhu2023towards}. We train five transformer models on AudioSet Strong labels and aggregate their predictions. We then consider the \textit{same} five models as students to be trained with KD on the aggregated predictions.
%
In this KD setting, we are free to modify the teachers since this will not impact the architecture of the students. We found it beneficial to increase each teacher's capacity with a large sequence model.

We apply this pre-training pipeline to five transformers and make the pre-trained weights publicly available.\footnote{\url{https://github.com/fschmid56/PretrainedSED}} The results of the temporally fine-grained PSDS1 metric~\cite{psds,Ebbers2022psds} on the AudioSet Strong evaluation set show a big performance gain compared to prior work~\cite{li24atst-frame}.
%
We also verify that, for all five transformers, the added pre-training step consistently improves in-domain downstream tasks involving frame-level sound event predictions.

\begin{figure}[t!]
    \centering
    \includegraphics[width=0.7\linewidth]{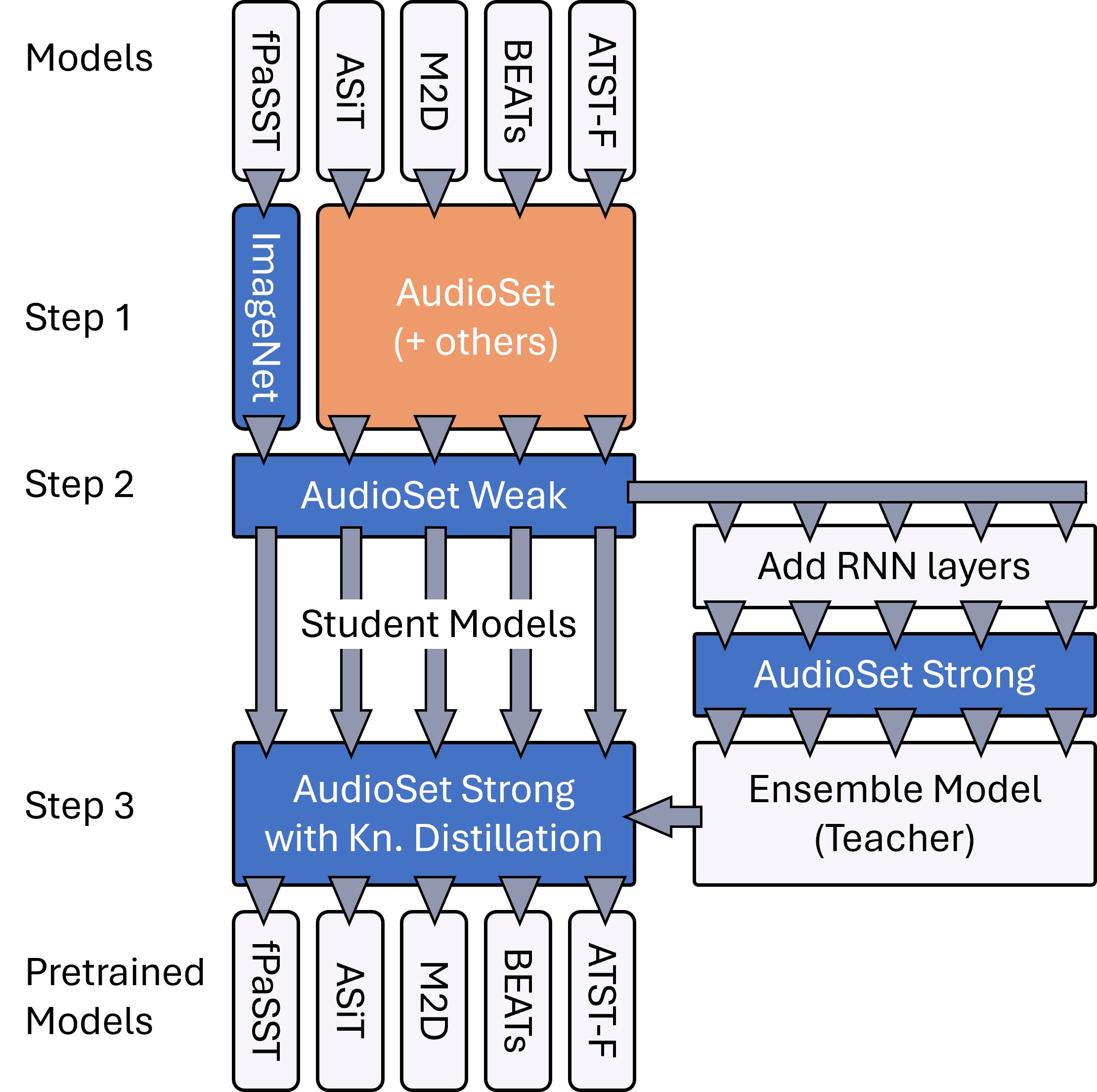}
    \caption{Full pre-training pipeline from scratch. Blue blocks stand for supervised training on the specified datasets and orange for self-supervised learning. }
    \label{fig:pipeline}
\end{figure}


\section{Proposed Pipeline}
Our pre-training pipeline can be articulated in three steps (see Fig.~\ref{fig:pipeline}). The first two steps are commonly performed in related work, while the third is our proposed extension, dedicated to frame-level downstream tasks.
In this section, we first outline how we adapt the model architecture for deriving frame-level predictions; then, we detail how the different model parts are trained in our proposed three-step pipeline.


\subsection{Model Architecture}
\label{subsec:model_comp}

We select five audio spectrogram transformers: fPaSST~\cite{schmid2024multi}, which was used by the top-ranked approach of Task 4 in the DCASE'24 Challenge~\cite{cornell2024dcase}; M2D~\cite{Niizumi24m2d-clap}, BEATs~\cite{Chen23BEATs} and ATST-Frame (ATST-F)~\cite{li24atst-frame}, which all showed strong performance on AudioSet Strong in~\cite{li24atst-frame}; and ASiT~\cite{Atito22ASiT}, a recently introduced transformer designed to learn local representations. 

The structure of all models trained on AudioSet Strong and the frame-level downstream tasks is defined as follows. Let us consider $g$, which represents each pre-trained transformer, stripped of the last layers dedicated to AudioSet Weak, ImageNet, or SSL. The model takes a mel spectrogram $\{\mathbf{x}\}_{t=1}^{T}$ with $T$ frames as input and outputs a sequence of embeddings of a certain length $S$ and dimension $D$.
\begin{equation}
\label{eq:forward}
    \{\mathbf{\hat{z}}_D\}_{t=1}^{S}   =   g(\{\mathbf{x}\}_{t=1}^{T}) 
\end{equation}

We stick to the default configurations for audio processing and model architectures proposed in the respective papers, leading to sequences with output dimensions of $S=250,496,250,62,497$ and $D=768,768,768,3840,768$ for ATST-F, BEATs, fPaSST, M2D, and ASiT for a 10-second audio clip, respectively. 


The output $\mathbf{\hat{z}}$ is then temporally aligned to a resolution of 40\,ms (corresponding to $S=250$ for 10-second audio clips) by using either adaptive average pooling (for $S>250$) or linear interpolation (for $S<250$). This enables the creation of the model ensemble for KD.

\begin{equation}
\label{eq:resampling}
\{\mathbf{\hat{e}}_D\}_{t=1}^{250} =  \underset{S \rightarrow 250}{\text{resample}}\big(\{\mathbf{\hat{z}}_D\}_{t=1}^{S}\big) \\
\end{equation}

\noindent
Finally, predictions are computed from the embeddings as:
\begin{equation}
\label{eq:final}
 \{\mathbf{\hat{o}}_C\}_{t=1}^{250} = \mathbf{W} f (\{\mathbf{\hat{e}}_D\}_{t=1}^{250}) + \mathbf{b} 
\end{equation}
\noindent
where $f$ denotes an optional sequence model that can be applied on top of the pre-trained model $g$. Its output is converted by a position-wise linear layer (parameterized by $\mathbf{W}$ and $\mathbf{b}$) into frame-level predictions $\{\mathbf{\hat{o}}_C\}_{t=1}^{250}$ with a dimension $C$ depending on the number of classes in the task (e.g., 447 for the training set of AudioSet Strong).

\subsection{Step 1: Self-Supervised / ImageNet Training}
\label{subsec:step1}


There are two common directions for the first pre-training step of many high-performance transformers. The first is to leverage the ImageNet dataset~\cite{Gong21Ast,Koutini21Passt} by loading ViT~\cite{Dosovitskiy20Image} or DeiT~\cite{Touvron21Deit} weights.
Although this is mostly popular among clip-level approaches, the DeiT-based fPaSST transformer in our selection has shown good performance for frame-level Sound Event Detection tasks in \cite{schmid2024multi}.

The second direction is to perform SSL on AudioSet audio data and optional complementary datasets~\cite{Dinkel24CED,Chen23BEATs,Niizumi24M2D,Niizumi24m2d-clap,Huang22Masked,Atito22ASiT,Gong2022ssast}. Different variants of this process are used in our models: ATST-F, BEATs, M2D, and ASiT. In contrast to ImageNet pre-training, this is performed on audio data, and involves patch-wise predictions, which is beneficial for learning local representations for frame-level downstream tasks.


\subsection{Step 2: Training on AudioSet Weak}
\label{subsec:step2}

After Step 1, we perform supervised training on AudioSet Weak. Our experiments in Sections~\ref{sec:as_exp} and \ref{sec:as_ds} show that this step improves the performance on frame-level tasks. 

Our pipeline builds on public pre-trained checkpoints.\footnote{External checkpoints are taken from the official GitHub repositories of the respective papers. Specific links are provided in our GitHub repository.}
ATST-F, BEATs, and M2D have checkpoints pre-trained on AudioSet Weak. For fPaSST and ASiT, we use the available checkpoints after Step 1,
then train on AudioSet Weak using the routine outlined in~\cite{Schmid22Efficient}, achieving a mean average precision of 48.4 and 49.8, respectively.

\subsection{Step 3: AudioSet Strong Knowledge Distillation}
\label{subsec:step3}
Step 3 is our main contribution to the pipeline, and it starts by adapting the transformers as outlined in Section~\ref{subsec:model_comp}. 
This step is based on two findings: 1) adding a sequence model $f$ that is trained from scratch on top of the transformer improves performance on AudioSet Strong, and 2) using ensemble KD allows for removing the additional complexity added by $f$, while further increasing single-model performance.

Regarding the choice of $f$, we tested different options, including multiple variants of transformer blocks, and we found that using a two-layer RNN with large hidden dimensions (1024 for fPaSST and 2048 for the others) substantially increases the predictive performance on AudioSet Strong. 

We build a well-performing ensemble (47.1 PSDS1 on AudioSet Strong) of 15 such models (3 per transformer type) by training each model independently with standard supervised cross-entropy loss, then averaging their logits $\{\mathbf{\hat{o}}_C\}_{t=1}^{250}$ class-wise for each sample's frame $t$ in the AudioSet Strong training set. The student models indicated in Fig.~\ref{fig:pipeline} are then trained without the additional complexity of a sequence model ($f$ is the identity function), but using KD to match the ensemble's predictions. In particular, we use the procedure outlined in~\cite{Schmid22Efficient}, which combines the supervised loss with a distillation loss, but we adapt it to compute the losses on frame level instead of clip level. 

\section{Experiments on AudioSet Strong}
\label{sec:as_exp}


This section details the training setup for Step 3 in our pipeline and presents results on the evaluation set of AudioSet Strong~\cite{audioset-strong}. We investigate how the performances of our models compare with existing work, and the impact of our design choices, such as the pre-training on AudioSet Weak, the usage of KD, and the additional RNN for teachers.

AudioSet Strong comes with pre-defined training and evaluation sets composed of 103,463 and 16,996 10-second audio clips, respectively. We successfully downloaded 100,911 clips from the training set and 16,935 clips from the evaluation set. We train on all 447 available sound classes in the training set and, in line with~\cite{li24atst-frame}, we evaluate against the 407 classes overlapping between training and evaluation sets. For comparison, we reproduce the pipeline of~\cite{li24atst-frame} since, in line with our work, they also focus on evaluation with high temporal resolution.



\subsection{Experimental Settings}
\label{subsec:exp_settings_as_strong}

We share the same training setup across all five transformers and only make a difference in training configurations for teachers and students. For both, we use the Adam optimizer~\cite{adam}, a cosine learning rate schedule with 5,000 warmup steps, and search for the best learning rate in a pre-defined grid (\{7e-5, 1e-4, 3e-4, 6e-4, 1e-3, 3e-3\}). The batch size is set to 256, and for postprocessing, we use a median filter with a size of 0.48 seconds, shared across all classes. Every experiment is executed three times with different seeds, and the results are averaged. 
We compute the statistical significance across different settings of our models by employing the ASO method~\cite{significance} with a confidence level of $\alpha = 0.05$, adjusted for the 5 model comparisons using the Bonferroni correction, and $\epsilon_\text{min} = 0.2$.

\paragraph{Teacher Models} 

As the size of AudioSet Strong ($\sim$100k) is comparably small to the size of AudioSet Weak ($\sim$2M), a key aspect to avoid overfitting and achieve high performance is aggressive data augmentation. In this regard, we apply frequency warping~\cite{li24atst-frame}, filter augmentation~\cite{Nam22filteraug}, Freq-MixStyle~\cite{Kim22fms,Schmid22dcase_ws}, and mixup~\cite{Zhang18mixup} on waveform and spectrogram levels. Furthermore, we borrow the idea of importance sampling~\cite{Kong20PANNs, Koutini21Passt}, a technique commonly used to balance classes when training on AudioSet Weak. Specifically, we compute sampling weights for each audio clip proportional to the inverse frequency of its labels, where label frequency is based on the total active time of a label in the training set.


\paragraph{Student Models}

Compared to the teachers, the students are trained to match the dense ensemble predictions in addition to the typical one-hot encoded labels. This additional complex target allows training the students in a simple setup using only mixup~\cite{Zhang18mixup}, uniform sampling, and training for an extended number of epochs (120 compared to 30 used for the teachers), without overfitting on the training data. For mixup, we not only mix spectrograms and the one-hot encoded labels but also the ensemble predictions accordingly.



\subsection{Metrics}
We use the threshold-independent PSDS1 metric \cite{Ebbers2022psds} that measures the intersection between ground truth events and detected events, focusing on the low reaction time (i.e., accurate localization of sound events), which perfectly aligns with our main interest. Other papers (e.g., \cite{audioset-strong,xie2023enhance}) use a metric that has a 960 ms resolution, which, in agreement with \cite{li24atst-frame}, we deem too coarse. Also, in agreement with this paper, we do not consider the variance penalty on the metric since it was designed for datasets with fewer and less imbalanced classes. Note that \cite{li24atst-frame} used the earlier threshold-dependent PSDS1, and we reevaluated their pipeline with the improved metric~\cite{Ebbers2022psds}.

\subsection{Results}
\label{subsec:as_strong_results}
Table~\ref{tab:audioset_strong} lists the results for our full proposed pipeline, without using KD in Step 3, without using RNNs for teacher models, and skipping the AudioSet Weak training (Step 2), and compares our results to related work~\cite{li24atst-frame}. 
The large performance gap to \cite{li24atst-frame} is statistically significant for all models and can be attributed to the combination of the three design choices ablated singularly in Table~\ref{tab:audioset_strong}, but also to the improved training routine on AudioSet Strong, i.e., the balanced sampling, and data augmentation as discussed in Section~\ref{subsec:exp_settings_as_strong}.

The strongest component in our pipeline is the ensemble KD procedure, which yields a statistically significant performance gain of at least 5\% for each transformer. Adding the RNN to the teacher model consistently improves the metric average, with statistical significance for BEATs and M2D. Pre-training on AudioSet Weak (Step 2) improves or matches performance across all models, with statistically significant gains for BEATs.
The performance gains from RNNs and Step~2 are smaller than we expected, given that we obtained on average 8.9\% and 7.7\% improvements when integrating these upgrades to single models.
These strong gains are less prominent when combining the single models to teacher ensembles, which reach PSDS1 scores of 47.1, 46.2, and 46.5 for the \textit{Proposed pipeline}, \textit{without RNN}, and \textit{without Step 2} settings, respectively.
As both components either improve or maintain performance without adding complexity to the final models, their inclusion in our pipeline is justified.

Overall, these experiments show that our pipeline can produce transformers with excellent fine-grained sound event detection capabilities that can be used directly if the desired target classes are represented in the AudioSet Strong class ontology.

\begin{table}[t!]

\centering
\caption{PSDS1 (without variance penalty) on AudioSet Strong.}
\centering 
\begin{tabular}{@{}lccccc@{}}
\toprule
 & ATST-F & BEATs  & fPaSST  & M2D & ASiT   \\ 
\midrule
Li et al.~\cite{li24atst-frame} & 40.9 & 36.5 & 38.7 & 36.9 & 37.0 \\
Proposed pipeline  & \textbf{45.8} & \textbf{46.5} & \textbf{45.4} & \textbf{46.3}  & \textbf{46.2}   \\
\hspace{1pt} --- without KD &  41.8 & 44.1 & 40.7  & 41.1  & 40.9  \\ 
\hspace{1pt} --- without RNN   & 45.7 & 45.8 & 45.3 & 46.0 & 46.1  \\
\hspace{1pt} --- without Step 2   & 45.7 & 46.3 & 45.2 & 44.9 & \textbf{46.2} \\ 
\bottomrule
\end{tabular}
\label{tab:audioset_strong}

\end{table}

\section{Experiments on Downstream Tasks}
\label{sec:as_ds}

This set of experiments evaluates our pre-trained models when fine-tuned on three downstream tasks. 
To study the contribution of every step in the pipeline, we compare our models after Step 1 (SSL or ImageNet), Step 2 (AudioSet Weak), and Step 3 (AudioSet Strong, i.e., full pipeline).


\subsection{Task Descriptions}
\label{subsec:task_desc}

We consider three frame-level downstream tasks: DCASE 2023 Task 4: \textit{Domestic Environment Sound Event Detection} (\textit{DESED})~\cite{desed} and two tasks from the HEAR benchmark~\cite{Turian21HEAR}: \textit{DCASE 2016 Task 2} (\textit{DC16-T2}) and \textit{MAESTRO 5hr} (\textit{MAESTRO}). For \textit{DESED}, we use the simplified training protocol from~\cite{li24atst-frame} that excludes unlabeled data from the training set, and for \textit{DC16-T2} and \textit{MAESTRO} we stick to the setup from~\cite{Turian21HEAR}. All three are multi-label classification tasks. 

\textit{DESED} and \textit{DC16-T2} are \textit{in-domain} tasks, as their labels describe domestic (e.g., ``Dishes'' or ``Frying'') and office-related (e.g., ``Coughing'' or ``Keyboard'') events, respectively, which aligns well with the AudioSet ontology~\cite{audioset2017Gemmeke}. In contrast, \textit{MAESTRO} is \textit{out-of-domain}, as it requires predicting the pitches of piano notes. In this regards, AudioSet Weak contains only a very limited amount of music-related labels, discerning some music genres and instruments. AudioSet Strong is even more simplified as it contains only one generic ``music'' label encompassing all musical events~\cite{audioset-strong}. 

\subsection{Experimental Settings}
We evaluate the pre-trained models in two configurations: \textit{Frozen} and \textit{Finetune}. In \textit{Frozen}, we freeze all model weights and only train a final task-specific linear layer. In \textit{Finetune}, the full model, including a task-specific linear layer, is updated. For finetuning on \textit{DESED}, aligned with \cite{Shao24atst-sed}, we use a layer-wise learning rate decay of 0.5 to avoid overfitting. As our goal is a fair comparison across the checkpoints, we opt for simple training routines and tune the learning rate separately with a grid search for each experiment and configuration. For \textit{MAESTRO}, we stick to the five-fold cross-validation used in~\cite{Turian21HEAR}, and for \textit{DESED} and \textit{DC16-T2}, every experiment is executed four times with different seeds.

To handle the longer clips (120 s) of \textit{MAESTRO} and \textit{DC16-T2}, we perform random 10-second crops during training and concatenate predictions on non-overlapping 10-second slices during evaluation. 



\subsection{Metrics}
For \textit{DESED}, we use the most commonly applied~\cite{cornell2024dcase} PSDS1 metric with variance penalty.
For the other two tasks, we use onset F-measure~\cite{Turian21HEAR}, but we reduce the tolerance for \textit{DC16-T2} from 200 ms to 50 ms (same as for \textit{MAESTRO}) since our interest is a temporally strict evaluation.

\subsection{Results}
\label{subsec:downstream_results}
The results in Table~\ref{tab:downstream} for the two in-domain tasks, i.e., \textit{DESED} and \textit{DC16-T2}, show that in all 20 experiments, the model trained with the full pipeline achieves the highest performance. We can conclude that pre-training with our pipeline consistently improves frame-level sound event detection tasks. As expected, the \textit{Finetune} configuration generally improves over the \textit{Frozen} setting. Since fPaSST is pre-trained on ImageNet in Step 1 without exposure to audio data, this results in the 0.0 values observed in its respective row.

Unfortunately, the results on the out-of-domain task, \textit{MAESTRO}, are not as conclusive, as the top results are spread across checkpoints from different pipeline steps. 
This hints at a limitation of our pipeline: substantial performance gains on downstream tasks can only be expected if audios and labels are related to the AudioSet ontology~\cite{audioset2017Gemmeke}. 


\begin{table}[t]
\centering
\caption{Evaluation on Downstream Tasks.}
\centering 
\begin{tabular}{@{}llcccccc@{}}
\toprule
& & \multicolumn{3}{c}{Frozen} & \multicolumn{3}{c}{Finetune} \\ 
\cmidrule(lr){3-5}
\cmidrule(lr){6-8}
\textit{Model} & \textit{Checkpoint} & \rotatebox{90}{\textit{DESED}} & \rotatebox{90}{\textit{DC16-T2}}  & \rotatebox{90}{\textit{MAESTRO}} & \rotatebox{90}{\textit{DESED}} & \rotatebox{90}{\textit{DC16 T2}}  & \rotatebox{90}{\textit{MAESTRO}}  \\ 
\midrule
\multirow{3}{*}{ATST-F} & Step 1 & 27.0 & 75.7 & 4.6 & 41.4 & 91.0 &  50.0 \\
 & Step 2 & 31.8 & 65.3 & 7.6 & 42.8 & 92.5 & 52.9 \\
 & Full Pipeline & \textbf{47.7} & \textbf{77.3} &  \textbf{21.9} & \textbf{48.2} & \textbf{92.9} & \textbf{55.0} \\ 
 \midrule
\multirow{3}{*}{BEATs} & Step 1 & 14.0 & 48.3 &  \textbf{24.4} & 35.0 & 76.5  & 26.1 \\
 & Step 2 & 36.0 & 51.5 & 12.6 & 37.3 & 77.4 & 38.5 \\
 & Full Pipeline & \textbf{48.1} & \textbf{58.3} & 0.3 & \textbf{49.2} & \textbf{84.7} & \textbf{51.2} \\
 \midrule
\multirow{3}{*}{fPaSST} & Step 1 & 0.0 & 0.0 &  0.0 & 0.0 & 82.9 & \textbf{41.9} \\
 & Step 2 & 40.1 & 58.0 & \textbf{43.5} & 39.1 & 88.3 & \textbf{41.9} \\
 & Full Pipeline & \textbf{45.4} & \textbf{72.0} & 28.7 & \textbf{48.2} & \textbf{88.8} & 38.5 \\
 \midrule
\multirow{3}{*}{M2D} & Step 1 & 27.7 & 69.3 & \textbf{30.3} & 38.2 & 81.9 & 35.9 \\
 & Step 2 & 39.9 & 67.4 & 25.1 & 40.6 & 84.4 & \textbf{36.6} \\
 & Full Pipeline & \textbf{49.2} & \textbf{70.1} & 22.3 & \textbf{48.7} & \textbf{84.8}  & 34.9 \\
 \midrule
\multirow{3}{*}{ASiT} & Step1 & 9.5 & 23.7 & 4.6 & 17.4 & 82.4 & 41.6 \\
 & Step 2 & 31.2 & 42.6 & \textbf{5.8} & 29.8 & 83.6 & 46.5 \\
 & Full Pipeline & \textbf{48.1} & \textbf{62.9} & 4.2 & \textbf{47.6} & \textbf{84.1} & \textbf{47.2} \\
 \bottomrule
\end{tabular}
\label{tab:downstream}
\end{table}




\section{Conclusion}

We presented a pre-training pipeline for transformers targeting sound event detection tasks. While prior research used the frame-level annotations of AudioSet, we are the first to leverage them as pre-training data for improving downstream frame-level tasks. Specifically, we use a balanced sampler, an aggressive data augmentation scheme, and ensemble knowledge distillation to effectively train single models on AudioSet frame-level labels. For five transformers, we show that this additional pre-training step benefits downstream in-domain sound event detection tasks. In addition to the pre-training guidelines, this research provides a main concrete outcome: public pre-trained checkpoints for five transformers capable of generating temporally fine-grained audio embeddings. We release these resources to aid the community in advancing audio classification tasks
and hope our research will also benefit other tasks relying on pre-trained audio encoders, such as audio captioning, text-to-audio grounding, or audio retrieval.

\bibliographystyle{IEEEtran}
\bibliography{refs}

\end{document}